\documentclass[12pt]{elsarticle}
\usepackage{epsfig,color}

\usepackage{amsmath}
\usepackage{amssymb}
\usepackage{graphicx}

\journal{PNFA - invited paper for proc. of TACONA 2012}

\begin{document}

\begin{frontmatter}

\author{N. Asger Mortensen}
\ead{asger@mailaps.org}
\address{DTU Fotonik, Department of Photonics Engineering, Technical University of Denmark, DK-2800 Kongens Lyngby, Denmark\\Center for Nanostructured Graphene (CNG), Technical University of Denmark, DK-2800 Kgs. Lyngby, Denmark
}

\title{Nonlocal formalism for nanoplasmonics: phenomenological and semi-classical considerations}

\begin{abstract}
  The plasmon response of metallic nanostructures is anticipated to exhibit nonlocal dynamics of the electron gas when exploring the true nanoscale. We extend the local-response approximation (based on Ohm's law) to account for a general short-range nonlocal response of the homogeneous electron gas. Without specifying further details of the underlying physical mechanism we show how this leads to a Laplacian correction term in the electromagnetic wave equation. Within the hydrodynamic model we demonstrate this explicitly and we identify the characteristic nonlocal range to be $\xi_{\scriptscriptstyle\rm NL}\sim v_F/\omega$ where $v_F$ is the Fermi velocity and $\omega$ is the optical angular frequency. For noble metals this gives significant corrections when characteristic device dimensions approach $\sim$1--10\,nm, whereas at more macroscopic length scales plasmonic phenomena are well accounted for by the local Drude response.
\end{abstract}


\end{frontmatter}

\section{Introduction}

The interaction of light with the free electrons in noble metals has led to a range of novel plasmonic phenomena and a versatile platform for a variety of new applications~\cite{Lal:2007,Schuller:2010,Atwater:2010,Brolo:2012,Kauranen:2012}. In particular, nanofabrication technologies and chemical synthesis are now allowing the plasmonics community to explore and manipulate light-matter interactions at sub-wavelength length scales, taking advantage of the spatially rapid oscillations of surface-plasmon polaritons and their ability to localize their energy in very small metallic volumes and structures~\cite{Gramotnev:2010}.

The understanding of the optical response of plasmonic structures has been successfully developed within the common framework of the local-response approximation (LRA) with Ohm's law ${\mathbf J}({\bf r}) = \sigma({\bf r}) {\mathbf E}({\bf r})$ as the constitutive equation~\cite{Maier:2007,Stockman:2011a}. However, the ability to fabricate and experimentally explore yet smaller metallic nanostructures has recently stimulated new theoretical developments aiming at quantum phenomena in nanoplasmonic systems~\cite{Jacob:2011,vanHulst:2012,Tame:2013} and the most recent experimental developments~\cite{Scholl:2012,Ciraci:2012,Savage:2012,Raza:2013,Scholl:2013} have clearly made a call for theory developments going beyond the LRA. Spatial dispersion due to nonlocal response is one of the extensions of the LRA formalism which have been studied extensively in more recent years~\cite{Fuchs:1987,Garcia-de-Abajo:2008,Aizpurua:2008,McMahon:2009,Raza:2011a,David:2011,Toscano:2012a,Toscano:2012b,Fernandez-Dominguez:2012,Wiener:2012,Huang:2013}. While the commonly employed LRA is inherently a description without any intrinsic length scales~\cite{Maier:2007,Stockman:2011a}, the new developments naturally introduce fundamental length scales associated with the quantum wave dynamics of the electron gas. As a consequence, plasmon polaritons can not sustain spatial oscillations beyond a cut-off wave number $\omega/v_F$~\cite{Yan:2012,Raza:2013b}, where $v_F$ is the Fermi velocity of the electron gas. However, even before reaching this cutoff we anticipate important nonlocal corrections in noble-metal nanostructures with characteristic dimension approaching the 1--10 nanometer regime.

For light interaction with arbitrarily shaped plasmonic structures, we begin from general considerations of nonlocal response and treat the case of short-range corrections to the LRA. The main result of this phenomenological analysis is that, irrespectively of the detailed underlying physical mechanism, nonlocal corrections appear in Maxwell's wave equation for the electrical field through an additional Laplacian operator term. Next, we turn to a specific hydrodynamic model and derive this result explicitly. Finally, we briefly address the importance of nonlocal response by dimensional analysis.

\section{Nonlocal response formalism}
We consider the interaction of light with metallic nanostructures in the linear regime, but with a general and spatially nonlocal response, i.e.
\begin{equation}\label{eq:nonlocal}
{\mathbf\nabla}\times{\mathbf\nabla}\times{\mathbf E}({\bf r})=\left(\tfrac{\omega}{c}\right)^2\int d{\bf r}'\, {\mathbf \varepsilon}({\bf r},{\bf r}') {\mathbf E}({\bf r}').
\end{equation}
In the following we will focus on the electron plasma itself, for simplicity leaving out any interband effects. For comparison to the commonly employed framework, we note that within the LRA the two-point dielectric function simplifies to ${\mathbf \varepsilon}({\bf r},{\bf r}') \approx \varepsilon_{\scriptscriptstyle D}{\mathbf \delta}({\bf r}-{\bf r}')$ with the usual Drude dielectric function
\begin{equation}
\varepsilon_{\scriptscriptstyle D}=1+i\sigma/(\varepsilon_0\omega)=1-\omega_p^2/[\omega(\omega+i/\tau)]
\end{equation}
where $\omega_p$ is the plasma frequency, $\sigma$ is the Ohmic conductivity, and $1/\tau$ is the damping rate. In this case, the integral in the integro-differential equation [Eq.~(\ref{eq:nonlocal})] is readily performed and we arrive at the ordinary partial-differential equation (PDE) for the local-response dynamics of plasmonic systems.

In the following we will use different approaches to get more insight into the nonlocal response function ${\mathbf \varepsilon}({\bf r},{\bf r}')$ associated with the plasmon response of the electron gas in metals.

\section{Phenomenological considerations}

First, we note that the local approximation with a delta-function response is an overall very good approximation and modeling based on this is indeed offering very good accounts of the majority of plasmonic phenomena observed in experiments. Thus, in our attempt to account for nonlocal response it seems adequate to only slightly relax the delta-function response. Consequently, we turn to a general nonlocal response function $\varepsilon({\bf r},{\bf r}')$ which is only short-range and with a characteristic nonlocal length $\xi_{\scriptscriptstyle\rm NL}$, such as in a Gaussian representation of a delta-function response (see Fig.~\ref{fig2}). For convenience, we write the response function as

\begin{equation}\label{eq:epsilon-f}
\varepsilon({\bf r},{\bf r}')=\varepsilon_{\scriptscriptstyle D}\delta({\bf r}-{\bf r}')+f(|{\bf r}-{\bf r}'|),
\end{equation}
i.e. with a local-response Drude contribution and with a small nonlocal correction associated with a homogeneous and isotropic plasma. In accordance with the above discussion we assume that $f$ satisfies
\begin{subequations}
\begin{equation}
\label{eq:f0}
\int d{\bf r}\,f(r)\ll \big|\varepsilon_{\scriptscriptstyle D}\big|,
\end{equation}
\begin{equation}
\label{eq:f1}
\int d{\bf r}\,rf(r)=0,
\end{equation}
\begin{equation}
\label{eq:f2}
\int d{\bf r}\,r^2f(r)=\xi_{\scriptscriptstyle\rm NL}^2.
\end{equation}
\end{subequations}
The approach in Eq.~(\ref{eq:epsilon-f}) is strongly inspired by a recent phenomenological approach by Ginzburg and Zayats~\cite{Ginzburg:2013a}. Rather than using a particular $f$ as a smearing function in numerical simulations we here proceed analytically. Due to the short-range behavior of $f$ we may conveniently Taylor expand the slowly varying electrical field in the integrand of Eq.~(\ref{eq:nonlocal}) around the point ${\bf r}$. To second order in $({\bf r}'-{\bf r})$ this gives
\begin{equation}
{E}_j({\bf r}')\simeq {E}_j({\bf r})+\left[{\mathbf \nabla} {E}_j({\bf r})\right]\cdot[{\bf r}'-{\bf r}]\\+[{\bf r}'-{\bf r}]^T \left[\hat{\mathbf H} {E}_j({\bf r})\right][{\bf r}'-{\bf r}],
\end{equation}
where the Hessian matrix $\hat{\mathbf H}$ has elements $\hat{H}_{ji}=\partial^2/(\partial_j\partial_i)$ and $j=x, y, z$. Next, substituting into Eq.~(\ref{eq:nonlocal}) and performing the integral it is clear that by assumption the zero-order term in the expansion contributes negligibly compared to the Drude contribution [Eq.~(\ref{eq:f0})]. For symmetry reasons, the first-order terms and the second-order cross terms vanish identically as they involve the first moment of $f$ [Eq.~(\ref{eq:f1})]. Consequently, the leading correction comes from the diagonal terms of the Hessian which we conveniently write as $\frac{1}{2} \left[\nabla^2 {E}_j({\bf r})\right][{\bf r}'-{\bf r}]^2$, i.e. involving the second moment of $f$ [Eq.~(\ref{eq:f2})]. With these steps we now get

\begin{equation}\label{eq:wave-laplacian-phenomenological}
{\mathbf\nabla}\times{\mathbf\nabla}\times{\mathbf E}({\bf r})=\left(\tfrac{\omega}{c}\right)^2\left[\varepsilon_{\scriptscriptstyle D}+{\mathcal C}_{\scriptscriptstyle\rm NL}\nabla^2\right]{\mathbf E}({\bf r}),
\end{equation}
where ${\mathcal C}_{\scriptscriptstyle\rm NL}=\tfrac{1}{2}\xi_{\scriptscriptstyle\rm NL}^2$. This is already a quite interesting result, suggesting that nonlocal corrections can be cast into a Laplacian term, seemingly independent on the microscopic origin of the nonlocal response. We also note that with the above few assumptions, we have formally turned from an explicit nonlocal integro-differential equation [Eq.~(\ref{eq:nonlocal})] to a PDE with the explicit appearance of only a single coordinate ${\bf r}$. Mathematically, the wave dynamics is now formally formulated as a local-response problem while it is clear that the nonlocal physics is still represented in the problem, i.e. by an explicit appearance of the Laplacian operator on the right-hand side. Clearly, the single-line form [Eq.~(\ref{eq:wave-laplacian-phenomenological})] is beneficial for the conceptual understanding and further analytical progress. From a numerical perspective, we note that the additional Laplacian does not add any complications beyond those
already posted by the double-curl operator on the left-hand side of the equation. Thus, effects of nonlocal response can now readily be included in existing simulation packages building on the local-response framework without significant computational cost.

At this stage, ${\mathcal C}_{\scriptscriptstyle\rm NL}$ is a coefficient of a purely phenomenological origin just as $\varepsilon_{\scriptscriptstyle D}$ accounts phenomenologically for the polarization in response to a local electrical field. In more detailed modeling, the semi-classical Drude model for $\varepsilon_{\scriptscriptstyle D}$ is commonly represented by tabulated experimental data. A similar strategy could in principle be pursued for the nonlocal coefficient ${\mathcal C}_{\scriptscriptstyle\rm NL}$ too. Alternatively, more microscopic or semi-classical approaches could be used to determine the coefficient. In the following we turn to a specific hydrodynamic model that accounts semi-classically for nonlocal response in a homogenous electron gas. As we shall see,  ${\mathcal C}_{\scriptscriptstyle\rm NL}\propto (v_F/\omega)^2$ and consequently the nonlocal response causes a smearing on length scales $\xi_{\scriptscriptstyle\rm NL}\sim v_F/\omega$.

\section{Hydrodynamic model}
For simplicity, we use a semi-classical hydrodynamic equation-of-motion for the jellium response of the homogeneous electron gas (originally put forward by Bloch~\cite{Bloch:1933a}), while restraining ourselves from the sub-nanometer regime where the electron gas is anticipated to exhibit its full quantum wave nature, including quantum tunneling (in dimer structures), quantum confinement effects (inhomogeneous electron densities in sub-nanometer particles with $a$ becoming comparable to $\lambda_F$), smearing of the surface charge-density profile (on the $\lambda_F$ scale due to a finite work function), and Friedel oscillations (on the $\lambda_F$ scale right inside the surfaces of metals).
While nonlocal response, or spatial dispersion, is a consequence of the quantum many-body properties of the electron gas, we here limit ourselves to a semi-classical treatment~\cite{Bloch:1933a,Barton:1979a,Boardman:1982a,Pitarke:2007a}. The usual equation-of-motion for an electron in an electrical field is extended to a hydrodynamic equation, which includes a pressure term that originates from the quantum kinetics of the electron gas. The equation-of-motion is in itself a nonlinear equation in several of the physical fields, but in the spirit of the linear-response form of Eq.~(\ref{eq:nonlocal}), equations are conveniently linearized, leading to the following set of coupled PDEs~\cite{Raza:2011a}
\begin{subequations}
\label{eq:coupledequations}
\begin{equation}
{\mathbf\nabla}\times{\mathbf\nabla}\times{\mathbf E}({\bf r})=\left(\tfrac{\omega}{c}\right)^2{\mathbf E}({\bf r}) +
i\omega\mu_0 {\mathbf J}({\bf r}),\label{eq:Maxwell}
\end{equation}
\begin{equation}
	\tfrac{\beta^2}{\omega\left(\omega+i/\tau\right)} {\mathbf \nabla} \left[ {\mathbf \nabla} \cdot {\mathbf J}({\bf r}) \right] +
 {\mathbf J}({\bf r}) = \sigma {\mathbf E}({\bf r}).
\label{eq:lmotion}
\end{equation}
\end{subequations}
Here, the ${\mathbf \nabla} \left[ {\mathbf \nabla} \cdot  {\mathbf J}\right]={\mathbf\nabla}\times{\mathbf\nabla}\times{\mathbf J}+ {\mathbf \nabla}^2 {\mathbf J}$ correction to Ohm's law has a strength $\beta^2/\omega^2$. The $\beta$ factor depends on the particular model for the electron gas and $\beta^2=(3/5) v_F^2$ within the Thomas--Fermi model.
In the local response-approximation the current field ${\mathbf J}$ can be straightforwardly eliminated and the resulting wave equation in the electrical field ${\mathbf E}$ needs only be solved subject to the usual boundary conditions for the electrical field ${\mathbf E}$ itself. On the other hand, the solution of the coupled equations calls for an additional boundary condition associated with  ${\mathbf J}$. We emphasize that once the physical context has been specified there is no freedom left for choosing the additional boundary condition; it follows explicitly from the already existing boundary conditions and the fulfilment of the continuity equation for the induced charge fluctuations~\cite{Raza:2011a,Yan:2012}. Neglecting the quantum leakage of electrons (hard-wall confinement associated with a very high workfunction), i.e. assuming that all electron wave functions vanish at the surface, the boundary conditions for the current at the metal surface are particular simple: no electrons can leave the metal volume, while they are allowed to move parallel to the surface. Mathematically, this implies that the tangential component of ${\mathbf J}$ is unrestricted while the normal component vanishes~\cite{Raza:2011a}. Inter-band effects can be included if the boundary conditions are modified accordingly, see the detailed discussion in the appendix of Ref.~\cite{Yan:2012}.

While $\mathbf k$-space representations of the hydrodynamic model have been conveniently applied to translationally invariant problems for decades~\cite{Barton:1979a,Boardman:1982a,Pitarke:2007a}, nonlocal response can now also be addressed in finite-size metallic structures of more arbitrary shape. In particular, the coupled equations [Eq.~(\ref{eq:coupledequations})] form a natural starting point for numerical solutions of arbitrarily shaped metallic nanostructures~\cite{Toscano:2012a,Hiremath:2012}. For numerical solutions we have employed a finite-element method~\cite{Toscano:2012a} and we have made our numerical implementation (an extension to the RF Module of COMSOL 4.2a) freely available~\cite{code}.

In order to illustrate the nonlocal nature more explicitly we next rewrite the hydrodynamic model so that it explicitly reflects the integro-differential form in Eq.~(\ref{eq:nonlocal}). Eliminating the current from Eq.~(\ref{eq:Maxwell}) we get
\begin{equation}\label{epsilon-hydrodynamic}
{\mathbf \varepsilon}({\bf r},{\bf r}')={\mathbf \delta}({\bf r}-{\bf r}')+\frac{i\sigma}{\varepsilon_0\omega}{\mathbf G}({\bf r},{\bf r}')
\end{equation}
with the dyadic Green's function defined by
\begin{equation}
\left\{	\tfrac{\beta^2}{\omega\left(\omega+i/\tau\right)} {\mathbf \nabla} \left[ {\mathbf \nabla} \cdot  \right] +
 1\right\} {\mathbf G}({\bf r},{\bf r}') =  {\mathbf \delta}({\bf r}-{\bf r}').
\end{equation}
The first term in Eq.~(\ref{epsilon-hydrodynamic}) is the local-response vacuum polarization while the second term is the nonlocal hydrodynamic contribution due to the jelly response of the electron gas. Obviously, ${\mathbf G}({\bf r},{\bf r}')$ has a short-range nature with a characteristic length scale $\beta/\omega$. Furthermore, for $\beta\rightarrow 0$ the dyadic Green's function approaches a Dirac delta function and we recover the usual LRA discussed below Eq.~(\ref{eq:nonlocal}).

Finally, let us now show how Eq.~(\ref{eq:coupledequations}) links up with Eq.~(\ref{eq:wave-laplacian-phenomenological}). Isolating the current in Eq.~(\ref{eq:Maxwell}) and substituting into Eq.~(\ref{eq:lmotion}) we get~\cite{Toscano:2013}

\begin{subequations}
\label{eq:operatorform}
\begin{equation}
{\mathbf\nabla}\times{\mathbf\nabla}\times{\mathbf E}({\bf r})=\left(\tfrac{\omega}{c}\right)^2 \hat\varepsilon_{\scriptscriptstyle\rm NL}({\bf r}){\mathbf E}({\bf r}),
\end{equation}
\begin{equation}
\hat\varepsilon_{\scriptscriptstyle\rm NL}({\bf r})=\varepsilon_{\scriptscriptstyle D}({\bf r})+\frac{\beta^2}{\omega(\omega+i/\tau)}{\bf \nabla}^2.
\end{equation}
\end{subequations}
Comparing to Eq.~(\ref{eq:wave-laplacian-phenomenological}) we recover the result obtained from more general considerations and we also infer that within the hydrodynamic model ${\mathcal C}_{\scriptscriptstyle\rm NL}=\frac{\beta^2}{\omega(\omega+i/\tau)}$, as anticipated in Sec.~3.

Equation~(\ref{eq:operatorform}) was recently used in our numerical study of waveguiding in plasmonic nanowires, grooves, and wedges~\cite{Toscano:2013} giving results in full agreement with those obtained numerically from Eq.~(\ref{eq:coupledequations}), while the former holds potential for better numerical convergence.

\section{Dimensional analysis}

\begin{figure}[t!]
\begin{center}
\includegraphics[width=0.5\columnwidth]{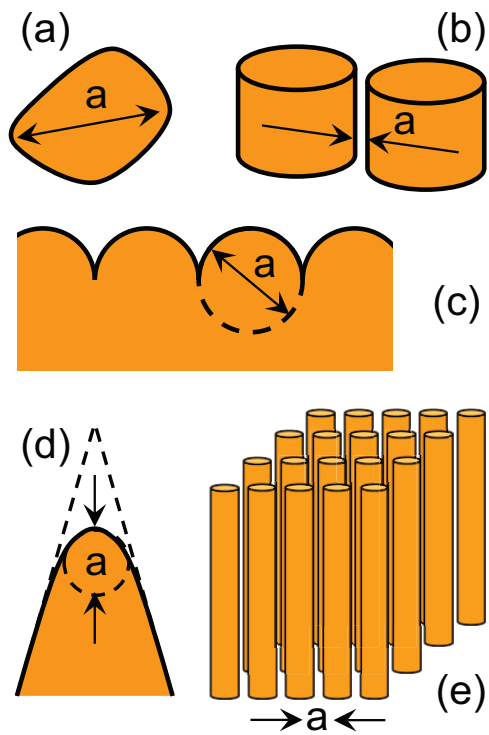}
\end{center}
\caption{Examples of metallic nanostructures, indicating characteristic length scales $a$ for (a) isolated metal particles, (b) metal-particle dimers with gaps, (c) corrugated metal surfaces, (d) sharp metal tips, and (e) metal-nanowire metamaterials. While the local-response approximation is typically adequate for $ka\gtrsim 1$, the nonlocal correction becomes important for $ka\ll 1$.}
\label{fig1}
\end{figure}

\begin{figure}[t!]
\begin{center}
\includegraphics[width=0.9\columnwidth]{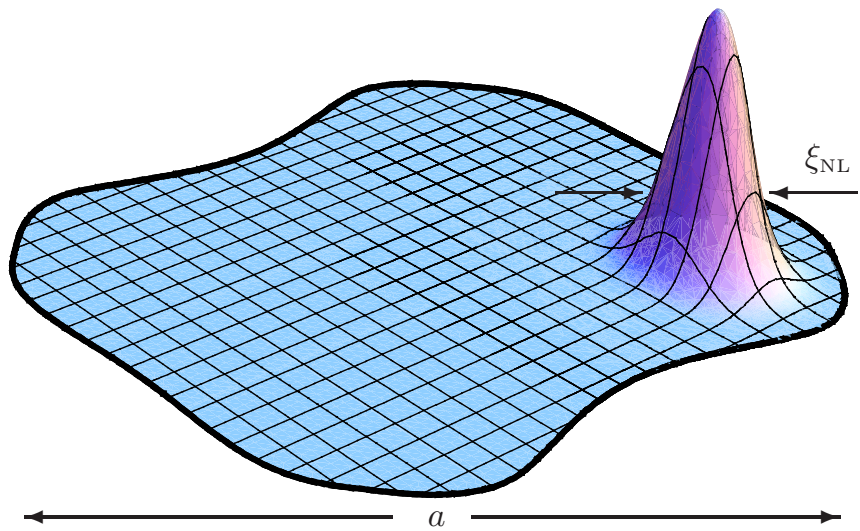}
\end{center}
\caption{Schematic illustration of an arbitrarily shaped metal domain with a characteristic dimension $a$. The nonlocal response function $f({\bf r},{\bf r}')$ with range $\xi_{\scriptscriptstyle\rm NL}$ is centered around ${\bf r}$ and plotted as a function of ${\bf r}'$. }
\label{fig2}
\end{figure}

\begin{figure}[t!]
\begin{center}
\includegraphics[width=0.9\columnwidth]{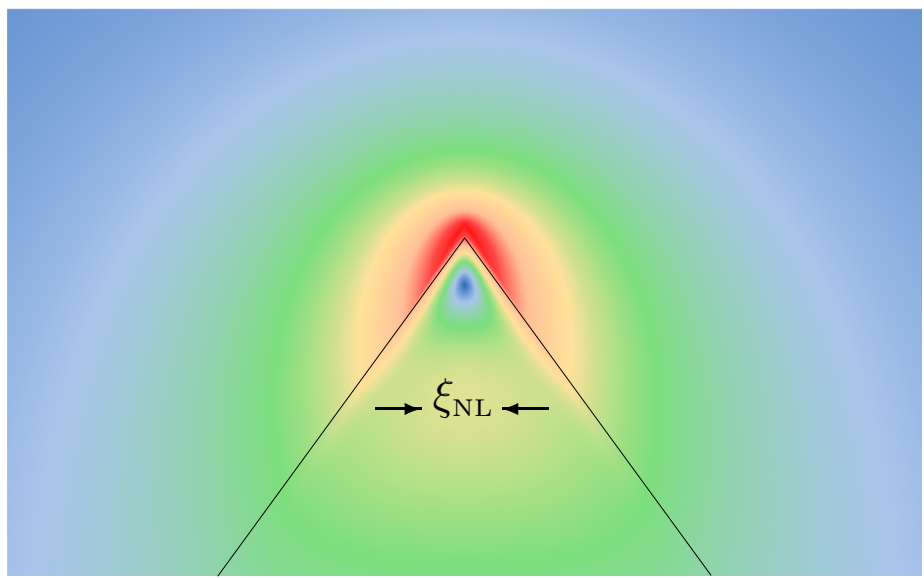}
\end{center}
\caption{Typical electrical-field intensity for the plane-wave electromagnetic excitation of an arbitrarily sharp metallic tip. The nonlocal response serves to smear out the diverging field of the otherwise singular geometry. Courtesy of G. Toscano~\cite{Toscano:thesis}.}
\label{fig3}
\end{figure}

Having determined  ${\mathcal C}_{\scriptscriptstyle\rm NL}$ in Eq.~(\ref{eq:wave-laplacian-phenomenological}) we now turn to a discussion of the importance of the nonlocal Laplacian correction. As expected, for $\beta\propto v_F \rightarrow 0$ we recover the usual LRA as discussed below Eq.~(\ref{eq:nonlocal}). Usually, this limit is considered trivially fulfilled because $v_F\ll c$ in most metals of relevance to plasmonics. However, from a dimensional analysis the correction to the local-response Drude contribution has a strength $(v_F/c)^2 (ka)^{-2}$ where $k=\omega/c$ is the free-space wave number while $a$ is a characteristic length scale, such as a nanoparticle diameter, a surface radius-of-curvature, or a gap distance separating two nearby metallic structures, see Fig.~\ref{fig1}. Thus, even though the electron dynamics is much slower than the speed of light the nonlocal response may nevertheless be an important correction when exhibited in structures with $ka\ll 1$. Fig.~\ref{fig3} shows a typical example where the metal surface has an abruptly varying surface topography. In the particular example, we show the electrical-field intensity distribution for the excitation of an arbitrarily sharp metallic tip. Clearly, the underlying geometry has a self-similarity thus possing a singular problem for wave equations without any intrinsic length scales. In the LRA such a structure would exhibit a strong field singularity whereas nonlocal response serves to smear the induced charges and the field intensity on the characteristic length scale $\xi_{\scriptscriptstyle\rm NL}\sim v_F/\omega$.

In order to estimate the spatial dimension where nonlocal response becomes a significant correction, say, $1\%$ relative to the Drude contribution we have $a\lesssim  10\times \xi_{\scriptscriptstyle\rm NL}$. Using $\omega_p$ as a characteristic frequency this can also be written as $a\lesssim  6\,$\AA$\times (r_s/a_0)^{1/2}$ where for the noble metals $r_s/a_0\sim 3$~\cite{Ashcroft:1976}. Consequently, the nonlocal response becomes an important correction only at the nanometer scale whereas at more macroscopic length scales plasmonic phenomena are well accounted for by the Drude response itself.

To further appreciate the importance of nonlocal response we briefly discuss its spectral influence on localized surface plasmon resonances (LSPR). Nonlocal hydrodynamic response is associated with additional dynamics of the electron gas and is thus in general causing a blueshift of the LSPR, see review by Kreibig and Genzel~\cite{Kreibig:1985} and references therein. As an example, the blueshift for a spherical metallic particle of diameter $D$ in a homogeneous dielectric background is
\begin{equation}
\Delta\omega \propto\frac{v_F}{D}+{\mathcal O}\left(\frac{1}{D^2}\right)
\label{eq:blueshift}
\end{equation}
where the prefactor is of the order unity, while depending on the dielectric background and the interband contributions to the metal polarizability~\cite{Raza:2013}. For silver particles, this seems in overall qualitative agreement with earlier experimental results on silver particles~\cite{Charle:1984,Charle:1989,Kreibig:1985} while our recent electron-energy-loss spectroscopy (EELS) on silver nanoparticles (supported by silicon nitride substrates) indicates a blueshift exceeding the above estimate~\cite{Raza:2013}.

\section{Discussion}

The present model generalizes the LRA to account for nonlocal response, while still assuming a homogeneous equilibrium density profile of the electron gas. Friedel oscillations and quantum leakage of the wave functions are examples of surface phenomena (on the scale of the Fermi wavelength) going beyond this latter assumption~\cite{Lang:1970}. These effects are always present at the surface no matter the size of the particle. If $D$ is further reduced, the equilibrium density in the bulk of the particle is influenced too, and eventually quantum-confinement effects with formation of discrete energy levels could be anticipated when $D$ approaches the Fermi wavelength of the electron gas itself. As emphasized in Ref.~\cite{Kreibig:1985} there is a number of further, sometimes competing, effects that can cause both redshifts and blueshifts in different metals as the metal-particle diameter is decreased below $10$\,nm.

Combining metallic nanoparticles in dimer and trimer configurations, large field enhancement can be supported by the dielectric gaps. The narrower the gap the larger field enhancement. Compared to the LRA, nonlocal response is smearing the induced charge profiles, thus constituting an important correction as the physical gap eventually turns comparable to $\xi_{\scriptscriptstyle\rm NL}$. In the context of the hybridization picture~\cite{Prodan:2003} the capacitor gap effectively appears wider and the coupling is consequently reduced. Thus, for the bonding mode with aligned dipoles the weakening of the field enhancement is associated with a blueshift as also confirmed by full numerical hydrodynamic simulations~\cite{Toscano:2012a}. If the gap is further reduced to the sub-nanometer scale where quantum leakage is likely to manifest itself, a short-circuiting of the capacitor is anticipated. This area is currently subject to much theoretical~\cite{Stella:2013,Teperik:2013,Andersen:2013} and experimental~\cite{Savage:2012,Scholl:2013} activity and it is stimulating phenomenological local~\cite{Esteban:2012} and nonlocal~\cite{Dong:2012} approaches trying to bridge the classical and the quantum mechanical regimes.

\section{Conclusion}

In conclusion, nonlocal response becomes noticeable in plasmonic structures with nanometer-sized features and characteristic dimensions. We have considered a short-range nonlocal response function for the homogeneous electron gas and shown how this after a few plausible assumptions leads to a Laplacian nonlocal correction term in the ordinary local-response wave equation, seemingly independent on the detailed underlying physical mechanism for the nonlocality. The result is in full agreement with a more detailed and specific semi-classical hydrodynamic model where the Laplacian term can be derived explicitly. Our work thus unifies different semi-classical models for nonlocal response and it provides an important link between our recent efforts within the hydrodynamic model~\cite{Toscano:2013} and the more phenomenological approach by Ginzburg and Zayats~\cite{Ginzburg:2013a}.

\section{Acknowledgments}

The 2009 paper by the Schatz group~\cite{McMahon:2009} stimulated a renewed interest in nonlocal response in arbitrarily shaped metallic nanostructures. During the efforts to establish the first consistent real-space formulation of the hydrodynamic model~\cite{Raza:2011a}, and in the subsequent explorations of this model, I have benefitted tremendously from enlightening discussions and stimulating feedback from colleagues, collaborators as well as present and former group members. In particular, I would like to acknowledge M. Wubs, A.-P.~Jauho, S.~I.~Bozhevolnyi, F.~J.~Garc{\'i}a de Abajo, P.~Nordlander, A.~I.~Fern{\'a}ndez-Dom{\'i}nguez, J.~Pendry, S.~A.~Maier, K.~S.~Thygesen, Hongxing~Xu, A. Zayats, T.~Garm~Pedersen, U.~Levy, S.~Xiao, W.~Yan. G.~Toscano, S.~Raza, T.~Christensen, J. Christensen, and N. Stenger.

The Center for Nanostructured Graphene is sponsored by the Danish National Research Foundation, Project DNRF58.


\begin{thebibliography}{49}
\providecommand{\natexlab}[1]{#1}
\providecommand{\url}[1]{\texttt{#1}}
\providecommand{\urlprefix}{URL }
\expandafter\ifx\csname urlstyle\endcsname\relax
  \providecommand{\doi}[1]{doi:\discretionary{}{}{}#1}\else
  \providecommand{\doi}[1]{doi:\discretionary{}{}{}\begingroup
  \urlstyle{rm}\url{#1}\endgroup}\fi
\providecommand{\bibinfo}[2]{#2}

\bibitem[{Lal et~al.(2007)Lal, Link, and Halas}]{Lal:2007}
\bibinfo{author}{S.~Lal}, \bibinfo{author}{S.~Link}, \bibinfo{author}{N.~J.
  Halas}, \bibinfo{title}{Nano-optics from sensing to waveguiding},
  \bibinfo{journal}{Nat. Photonics} \bibinfo{volume}{1}~(\bibinfo{number}{11})
  (\bibinfo{year}{2007}) \bibinfo{pages}{641--648}.

\bibitem[{Schuller et~al.(2010)Schuller, Barnard, Cai, Jun, White, and
  Brongersma}]{Schuller:2010}
\bibinfo{author}{J.~A. Schuller}, \bibinfo{author}{E.~S. Barnard},
  \bibinfo{author}{W.~Cai}, \bibinfo{author}{Y.~C. Jun}, \bibinfo{author}{J.~S.
  White}, \bibinfo{author}{M.~L. Brongersma}, \bibinfo{title}{Plasmonics for
  extreme light concentration and manipulation}, \bibinfo{journal}{Nat. Mater.}
  \bibinfo{volume}{9}~(\bibinfo{number}{3}) (\bibinfo{year}{2010})
  \bibinfo{pages}{193--204}.

\bibitem[{Atwater and Polman(2010)}]{Atwater:2010}
\bibinfo{author}{H.~A. Atwater}, \bibinfo{author}{A.~Polman},
  \bibinfo{title}{Plasmonics for improved photovoltaic devices},
  \bibinfo{journal}{Nat. Mater.} \bibinfo{volume}{9}~(\bibinfo{number}{3})
  (\bibinfo{year}{2010}) \bibinfo{pages}{205--213}.

\bibitem[{Brolo(2012)}]{Brolo:2012}
\bibinfo{author}{A.~G. Brolo}, \bibinfo{title}{Plasmonics for future
  biosensors}, \bibinfo{journal}{Nat. Photonics}
  \bibinfo{volume}{6}~(\bibinfo{number}{11}) (\bibinfo{year}{2012})
  \bibinfo{pages}{709--713}.

\bibitem[{Kauranen and Zayats(2012)}]{Kauranen:2012}
\bibinfo{author}{M.~Kauranen}, \bibinfo{author}{A.~V. Zayats},
  \bibinfo{title}{Nonlinear plasmonics}, \bibinfo{journal}{Nat. Photonics}
  \bibinfo{volume}{6}~(\bibinfo{number}{11}) (\bibinfo{year}{2012})
  \bibinfo{pages}{737--748}.

\bibitem[{Gramotnev and Bozhevolnyi(2010)}]{Gramotnev:2010}
\bibinfo{author}{D.~K. Gramotnev}, \bibinfo{author}{S.~I. Bozhevolnyi},
  \bibinfo{title}{Plasmonics beyond the diffraction limit},
  \bibinfo{journal}{Nat. Photonics} \bibinfo{volume}{4}~(\bibinfo{number}{2})
  (\bibinfo{year}{2010}) \bibinfo{pages}{83--91}.

\bibitem[{Maier(2007)}]{Maier:2007}
\bibinfo{author}{S.~A. Maier}, \bibinfo{title}{Plasmonics: Fundamentals and
  Applications}, \bibinfo{publisher}{Springer}, \bibinfo{address}{New York},
  \bibinfo{year}{2007}.

\bibitem[{Stockman(2011)}]{Stockman:2011a}
\bibinfo{author}{M.~I. Stockman}, \bibinfo{title}{Nanoplasmonics: past,
  present, and glimpse into future}, \bibinfo{journal}{Opt. Express}
  \bibinfo{volume}{19}~(\bibinfo{number}{22}) (\bibinfo{year}{2011})
  \bibinfo{pages}{22029--22106}.

\bibitem[{Jacob and Shalaev(2011)}]{Jacob:2011}
\bibinfo{author}{Z.~Jacob}, \bibinfo{author}{V.~M. Shalaev},
  \bibinfo{title}{Plasmonics Goes Quantum}, \bibinfo{journal}{Science}
  \bibinfo{volume}{334}~(\bibinfo{number}{6055}) (\bibinfo{year}{2011})
  \bibinfo{pages}{463--464}.

\bibitem[{van Hulst(2012)}]{vanHulst:2012}
\bibinfo{author}{N.~F. van Hulst}, \bibinfo{title}{Plasmon quantum limit
  exposed}, \bibinfo{journal}{Nat. Nanotechnol.}
  \bibinfo{volume}{7}~(\bibinfo{number}{12}) (\bibinfo{year}{2012})
  \bibinfo{pages}{775--777}.

\bibitem[{Tame et~al.(2013)Tame, McEnery, {\"O}zdemir, Lee, Maier, and
  Kim}]{Tame:2013}
\bibinfo{author}{M.~S. Tame}, \bibinfo{author}{K.~R. McEnery},
  \bibinfo{author}{S.~K. {\"O}zdemir}, \bibinfo{author}{J.~Lee},
  \bibinfo{author}{S.~A. Maier}, \bibinfo{author}{M.~S. Kim},
  \bibinfo{title}{Quantum plasmonics}, \bibinfo{journal}{Nat. Phys.}
  \bibinfo{volume}{9}~(\bibinfo{number}{6}) (\bibinfo{year}{2013})
  \bibinfo{pages}{329--340}.

\bibitem[{Scholl et~al.(2012)Scholl, Koh, and Dionne}]{Scholl:2012}
\bibinfo{author}{J.~A. Scholl}, \bibinfo{author}{A.~L. Koh},
  \bibinfo{author}{J.~A. Dionne}, \bibinfo{title}{Quantum plasmon resonances of
  individual metallic nanoparticles}, \bibinfo{journal}{Nature}
  \bibinfo{volume}{483}~(\bibinfo{number}{7390}) (\bibinfo{year}{2012})
  \bibinfo{pages}{421}.

\bibitem[{Cirac{\`i} et~al.(2012)Cirac{\`i}, Hill, Mock, Urzhumov,
  Fern{\'a}ndez-Dom{\'i}nguez, Maier, Pendry, Chilkoti, and
  Smith}]{Ciraci:2012}
\bibinfo{author}{C.~Cirac{\`i}}, \bibinfo{author}{R.~T. Hill},
  \bibinfo{author}{J.~J. Mock}, \bibinfo{author}{Y.~Urzhumov},
  \bibinfo{author}{A.~I. Fern{\'a}ndez-Dom{\'i}nguez}, \bibinfo{author}{S.~A.
  Maier}, \bibinfo{author}{J.~B. Pendry}, \bibinfo{author}{A.~Chilkoti},
  \bibinfo{author}{D.~R. Smith}, \bibinfo{title}{Probing the Ultimate Limits of
  Plasmonic Enhancement}, \bibinfo{journal}{Science}
  \bibinfo{volume}{337}~(\bibinfo{number}{6098}) (\bibinfo{year}{2012})
  \bibinfo{pages}{1072--1074}.

\bibitem[{Savage et~al.(2012)Savage, Hawkeye, Esteban, Borisov, Aizpurua, and
  Baumberg}]{Savage:2012}
\bibinfo{author}{K.~J. Savage}, \bibinfo{author}{M.~M. Hawkeye},
  \bibinfo{author}{R.~Esteban}, \bibinfo{author}{A.~G. Borisov},
  \bibinfo{author}{J.~Aizpurua}, \bibinfo{author}{J.~J. Baumberg},
  \bibinfo{title}{Revealing the quantum regime in tunnelling plasmonics},
  \bibinfo{journal}{Nature} \bibinfo{volume}{491}~(\bibinfo{number}{7425})
  (\bibinfo{year}{2012}) \bibinfo{pages}{574--577}.

\bibitem[{Raza et~al.(2013)Raza, Stenger, Kadkhodazadeh, Fischer, Kostesha,
  Jauho, Burrows, Wubs, and Mortensen}]{Raza:2013}
\bibinfo{author}{S.~Raza}, \bibinfo{author}{N.~Stenger},
  \bibinfo{author}{S.~Kadkhodazadeh}, \bibinfo{author}{S.~V. Fischer},
  \bibinfo{author}{N.~Kostesha}, \bibinfo{author}{A.-P. Jauho},
  \bibinfo{author}{A.~Burrows}, \bibinfo{author}{M.~Wubs},
  \bibinfo{author}{N.~A. Mortensen}, \bibinfo{title}{Blueshift of the surface
  plasmon resonance in silver nanoparticles studied with EELS},
  \bibinfo{journal}{Nanophotonics} \bibinfo{volume}{2}~(\bibinfo{number}{2})
  (\bibinfo{year}{2013}) \bibinfo{pages}{131--138}.

\bibitem[{Scholl et~al.(2013)Scholl, Garcia-Etxarri, Koh, and
  Dionne}]{Scholl:2013}
\bibinfo{author}{J.~A. Scholl}, \bibinfo{author}{A.~Garcia-Etxarri},
  \bibinfo{author}{A.~L. Koh}, \bibinfo{author}{J.~A. Dionne},
  \bibinfo{title}{{Observation of Quantum Tunneling between Two Plasmonic
  Nanoparticles}}, \bibinfo{journal}{Nano Lett.}
  \bibinfo{volume}{13}~(\bibinfo{number}{2}) (\bibinfo{year}{2013})
  \bibinfo{pages}{564--569}.

\bibitem[{Fuchs and Claro(1987)}]{Fuchs:1987}
\bibinfo{author}{R.~Fuchs}, \bibinfo{author}{F.~Claro},
  \bibinfo{title}{Multipolar response of small metallic spheres: Nonlocal
  theory}, \bibinfo{journal}{Phys. Rev. B}
  \bibinfo{volume}{35}~(\bibinfo{number}{8}) (\bibinfo{year}{1987})
  \bibinfo{pages}{3722--3727}.

\bibitem[{{Garc\'{i}a de Abajo}(2008)}]{Garcia-de-Abajo:2008}
\bibinfo{author}{F.~J. {Garc\'{i}a de Abajo}}, \bibinfo{title}{Nonlocal Effects
  in the Plasmons of Strongly Interacting Nanoparticles, Dimers, and
  Waveguides}, \bibinfo{journal}{J. Phys. Chem. C}
  \bibinfo{volume}{112}~(\bibinfo{number}{46}) (\bibinfo{year}{2008})
  \bibinfo{pages}{17983--17987}.

\bibitem[{Aizpurua and Rivacoba(2008)}]{Aizpurua:2008}
\bibinfo{author}{J.~Aizpurua}, \bibinfo{author}{A.~Rivacoba},
  \bibinfo{title}{Nonlocal effects in the plasmons of nanowires and
  nanocavities excited by fast electron beams}, \bibinfo{journal}{Phys. Rev. B}
  \bibinfo{volume}{78}~(\bibinfo{number}{3}) (\bibinfo{year}{2008})
  \bibinfo{pages}{035404}.

\bibitem[{McMahon et~al.(2009)McMahon, Gray, and Schatz}]{McMahon:2009}
\bibinfo{author}{J.~M. McMahon}, \bibinfo{author}{S.~K. Gray},
  \bibinfo{author}{G.~C. Schatz}, \bibinfo{title}{Nonlocal optical response of
  metal nanostructures with arbitrary shape}, \bibinfo{journal}{Phys. Rev.
  Lett.} \bibinfo{volume}{103}~(\bibinfo{number}{9}) (\bibinfo{year}{2009})
  \bibinfo{pages}{097403}.

\bibitem[{Raza et~al.(2011)Raza, Toscano, Jauho, Wubs, and
  Mortensen}]{Raza:2011a}
\bibinfo{author}{S.~Raza}, \bibinfo{author}{G.~Toscano}, \bibinfo{author}{A.-P.
  Jauho}, \bibinfo{author}{M.~Wubs}, \bibinfo{author}{N.~A. Mortensen},
  \bibinfo{title}{Unusual resonances in nanoplasmonic structures due to
  nonlocal response}, \bibinfo{journal}{Phys. Rev. B}
  \bibinfo{volume}{84}~(\bibinfo{number}{12}) (\bibinfo{year}{2011})
  \bibinfo{pages}{121412(R)}.

\bibitem[{David and Garc{\'i}a~de Abajo(2011)}]{David:2011}
\bibinfo{author}{C.~David}, \bibinfo{author}{F.~J. Garc{\'i}a~de Abajo},
  \bibinfo{title}{Spatial Nonlocality in the Optical Response of Metal
  Nanoparticles}, \bibinfo{journal}{J. Phys. Chem. C}
  \bibinfo{volume}{115}~(\bibinfo{number}{40}) (\bibinfo{year}{2011})
  \bibinfo{pages}{19470--19475}.

\bibitem[{Toscano et~al.(2012{\natexlab{a}})Toscano, Raza, Jauho, Mortensen,
  and Wubs}]{Toscano:2012a}
\bibinfo{author}{G.~Toscano}, \bibinfo{author}{S.~Raza}, \bibinfo{author}{A.-P.
  Jauho}, \bibinfo{author}{N.~A. Mortensen}, \bibinfo{author}{M.~Wubs},
  \bibinfo{title}{Modified field enhancement and extinction in plasmonic
  nanowire dimers due to nonlocal response}, \bibinfo{journal}{Opt. Express}
  \bibinfo{volume}{20}~(\bibinfo{number}{4})
  (\bibinfo{year}{2012}{\natexlab{a}}) \bibinfo{pages}{4176 -- 4188}.

\bibitem[{Toscano et~al.(2012{\natexlab{b}})Toscano, Raza, Xiao, Wubs, Jauho,
  Bozhevolnyi, and Mortensen}]{Toscano:2012b}
\bibinfo{author}{G.~Toscano}, \bibinfo{author}{S.~Raza},
  \bibinfo{author}{S.~Xiao}, \bibinfo{author}{M.~Wubs}, \bibinfo{author}{A.-P.
  Jauho}, \bibinfo{author}{S.~I. Bozhevolnyi}, \bibinfo{author}{N.~A.
  Mortensen}, \bibinfo{title}{Surface-enhanced Raman spectroscopy: nonlocal
  limitations}, \bibinfo{journal}{Opt. Lett.}
  \bibinfo{volume}{37}~(\bibinfo{number}{13})
  (\bibinfo{year}{2012}{\natexlab{b}}) \bibinfo{pages}{2538--2540}.

\bibitem[{Fern{\'a}ndez-Dom{\'i}nguez et~al.(2012)Fern{\'a}ndez-Dom{\'i}nguez,
  Wiener, Garc{\'i}a-Vidal, Maier, and Pendry}]{Fernandez-Dominguez:2012}
\bibinfo{author}{A.~I. Fern{\'a}ndez-Dom{\'i}nguez},
  \bibinfo{author}{A.~Wiener}, \bibinfo{author}{F.~J. Garc{\'i}a-Vidal},
  \bibinfo{author}{S.~A. Maier}, \bibinfo{author}{J.~B. Pendry},
  \bibinfo{title}{Transformation-Optics Description of Nonlocal Effects in
  Plasmonic Nanostructures}, \bibinfo{journal}{Phys. Rev. Lett.}
  \bibinfo{volume}{108}~(\bibinfo{number}{10}) (\bibinfo{year}{2012})
  \bibinfo{pages}{106802}.

\bibitem[{Wiener et~al.(2012)Wiener, Fern{\'a}ndez-Dom{\'i}nguez, Horsfield,
  Pendry, and Maier}]{Wiener:2012}
\bibinfo{author}{A.~Wiener}, \bibinfo{author}{A.~I.
  Fern{\'a}ndez-Dom{\'i}nguez}, \bibinfo{author}{A.~P. Horsfield},
  \bibinfo{author}{J.~B. Pendry}, \bibinfo{author}{S.~A. Maier},
  \bibinfo{title}{Nonlocal Effects in the Nanofocusing Performance of Plasmonic
  Tips}, \bibinfo{journal}{Nano Lett.}
  \bibinfo{volume}{12}~(\bibinfo{number}{6}) (\bibinfo{year}{2012})
  \bibinfo{pages}{3308--3314}.

\bibitem[{Huang et~al.(2013)Huang, Bao, and He}]{Huang:2013}
\bibinfo{author}{Q.~Huang}, \bibinfo{author}{F.~Bao}, \bibinfo{author}{S.~He},
  \bibinfo{title}{Nonlocal effects in a hybrid plasmonic waveguide for
  nanoscale confinement}, \bibinfo{journal}{Opt. Express}
  \bibinfo{volume}{21}~(\bibinfo{number}{2}) (\bibinfo{year}{2013})
  \bibinfo{pages}{1430--1439}.

\bibitem[{Yan et~al.(2012)Yan, Wubs, and Mortensen}]{Yan:2012}
\bibinfo{author}{W.~Yan}, \bibinfo{author}{M.~Wubs}, \bibinfo{author}{N.~A.
  Mortensen}, \bibinfo{title}{Hyperbolic metamaterials: Nonlocal response
  regularizes broadband supersingularity}, \bibinfo{journal}{Phys. Rev. B}
  \bibinfo{volume}{86}~(\bibinfo{number}{20}) (\bibinfo{year}{2012})
  \bibinfo{pages}{205429}.

\bibitem[{Raza et~al.(????)Raza, Christensen, Wubs, Bozhevolnyi, and
  Mortensen}]{Raza:2013b}
\bibinfo{author}{S.~Raza}, \bibinfo{author}{T.~Christensen},
  \bibinfo{author}{M.~Wubs}, \bibinfo{author}{S.~I. Bozhevolnyi},
  \bibinfo{author}{N.~A. Mortensen}, \bibinfo{title}{Nonlocal response in
  thin-film waveguides: loss versus nonlocality and breaking of
  complementarity}, \bibinfo{journal}{[arXiv:1305.1185]}.

\bibitem[{Ginzburg and Zayats(2013)}]{Ginzburg:2013a}
\bibinfo{author}{P.~Ginzburg}, \bibinfo{author}{A.~Zayats},
  \bibinfo{title}{Localized Surface Plasmon Resonances in Spatially Dispersive
  Nano-Objects: Phenomenological Treatise}, \bibinfo{journal}{ACS Nano}
  \bibinfo{volume}{7}~(\bibinfo{number}{5}) (\bibinfo{year}{2013})
  \bibinfo{pages}{4334–4342}.

\bibitem[{Bloch(1933)}]{Bloch:1933a}
\bibinfo{author}{F.~Bloch}, \bibinfo{title}{Bremsverm\text{\"{o}}gen von Atomen
  mit mehreren Elektronen}, \bibinfo{journal}{Z. Phys. A}
  \bibinfo{volume}{81}~(\bibinfo{number}{5-6}) (\bibinfo{year}{1933})
  \bibinfo{pages}{363--376}.

\bibitem[{Barton(1979)}]{Barton:1979a}
\bibinfo{author}{G.~Barton}, \bibinfo{title}{Some surface effects in the
  hydrodynamic model of metals}, \bibinfo{journal}{Rep. Prog. Phys.}
  \bibinfo{volume}{42}~(\bibinfo{number}{6}) (\bibinfo{year}{1979})
  \bibinfo{pages}{963--1016}.

\bibitem[{Boardman(1982)}]{Boardman:1982a}
\bibinfo{author}{A.~Boardman}, \bibinfo{title}{Electromagnetic Surface Modes.
  Hydrodynamic theory of plasmon-polaritons on plane surfaces.},
  \bibinfo{publisher}{John Wiley and Sons}, \bibinfo{year}{1982}.

\bibitem[{Pitarke et~al.(2007)Pitarke, Silkin, Chulkov, and
  Echenique}]{Pitarke:2007a}
\bibinfo{author}{J.~Pitarke}, \bibinfo{author}{V.~Silkin},
  \bibinfo{author}{E.~Chulkov}, \bibinfo{author}{P.~Echenique},
  \bibinfo{title}{Theory of surface plasmons and surface-plasmon polaritons},
  \bibinfo{journal}{Rep. Prog. Phys.}
  \bibinfo{volume}{70}~(\bibinfo{number}{1}) (\bibinfo{year}{2007})
  \bibinfo{pages}{1--87}.

\bibitem[{Hiremath et~al.(2012)Hiremath, Zschiedrich, and
  Schmidt}]{Hiremath:2012}
\bibinfo{author}{K.~R. Hiremath}, \bibinfo{author}{L.~Zschiedrich},
  \bibinfo{author}{F.~Schmidt}, \bibinfo{title}{Numerical solution of nonlocal
  hydrodynamic Drude model for arbitrary shaped nano-plasmonic structures using
  Nédélec finite elements}, \bibinfo{journal}{J. Comp. Phys.}
  \bibinfo{volume}{231}~(\bibinfo{number}{17}) (\bibinfo{year}{2012})
  \bibinfo{pages}{5890 -- 5896}.

\bibitem[{{NanoPlasmonics Lab}(2012)}]{code}
\bibinfo{author}{{NanoPlasmonics Lab}},
  \bibinfo{note}{\url{http://www.nanopl.org}}.

\bibitem[{Toscano et~al.(????)Toscano, Raza, Yan, Jeppesen, Xiao, Wubs, Jauho,
  Bozhevolnyi, and Mortensen}]{Toscano:2013}
\bibinfo{author}{G.~Toscano}, \bibinfo{author}{S.~Raza},
  \bibinfo{author}{W.~Yan}, \bibinfo{author}{C.~Jeppesen},
  \bibinfo{author}{S.~Xiao}, \bibinfo{author}{M.~Wubs}, \bibinfo{author}{A.-P.
  Jauho}, \bibinfo{author}{S.~I. Bozhevolnyi}, \bibinfo{author}{N.~A.
  Mortensen}, \bibinfo{title}{Nonlocal response in plasmonic waveguiding with
  extreme light confinement}, \bibinfo{journal}{[arXiv:1212.4925]}.

\bibitem[{Ashcroft and Mermin(1976)}]{Ashcroft:1976}
\bibinfo{author}{N.~W. Ashcroft}, \bibinfo{author}{N.~D. Mermin},
  \bibinfo{title}{Solid State Physics}, \bibinfo{publisher}{Saunders College
  Publishing}, \bibinfo{address}{Fort Worth}, \bibinfo{year}{1976}.

\bibitem[{Kreibig and Genzel(1985)}]{Kreibig:1985}
\bibinfo{author}{U.~Kreibig}, \bibinfo{author}{L.~Genzel},
  \bibinfo{title}{Optical absorption of small metallic particles},
  \bibinfo{journal}{Surf. Sci.} \bibinfo{volume}{156} (\bibinfo{year}{1985})
  \bibinfo{pages}{678--700}.

\bibitem[{{Charl\'{e}} et~al.(1984){Charl\'{e}}, Frank, and
  Schulze}]{Charle:1984}
\bibinfo{author}{K.-P. {Charl\'{e}}}, \bibinfo{author}{F.~Frank},
  \bibinfo{author}{W.~Schulze}, \bibinfo{title}{The Optical Properties of
  Silver Microcrystallites in Dependence on Size and the Influence of the
  Matrix Environment}, \bibinfo{journal}{Ber. Bunsen-Ges. Phys. Chem. Chem.
  Phys.} \bibinfo{volume}{88}~(\bibinfo{number}{4}) (\bibinfo{year}{1984})
  \bibinfo{pages}{350--354}.

\bibitem[{{Charl\'{e}} et~al.(1989){Charl\'{e}}, Schulze, and
  Winter}]{Charle:1989}
\bibinfo{author}{K.-P. {Charl\'{e}}}, \bibinfo{author}{W.~Schulze},
  \bibinfo{author}{B.~Winter}, \bibinfo{title}{The size dependent shift of the
  surface-plasmon absorption-band of small spherical metal particles},
  \bibinfo{journal}{Z. Phys. D} \bibinfo{volume}{12}~(\bibinfo{number}{1-4})
  (\bibinfo{year}{1989}) \bibinfo{pages}{471--475}.

\bibitem[{Lang and Kohn(1970)}]{Lang:1970}
\bibinfo{author}{N.~Lang}, \bibinfo{author}{W.~Kohn}, \bibinfo{title}{Theory of
  Metal Surfaces: Charge Density and Surface Energy}, \bibinfo{journal}{Phys.
  Rev. B} \bibinfo{volume}{1}~(\bibinfo{number}{12}) (\bibinfo{year}{1970})
  \bibinfo{pages}{4555--4568}.

\bibitem[{Prodan et~al.(2003)Prodan, Radloff, Halas, and
  Nordlander}]{Prodan:2003}
\bibinfo{author}{E.~Prodan}, \bibinfo{author}{C.~Radloff},
  \bibinfo{author}{N.~Halas}, \bibinfo{author}{P.~Nordlander},
  \bibinfo{title}{A hybridization model for the plasmon response of complex
  nanostructures}, \bibinfo{journal}{Science}
  \bibinfo{volume}{302}~(\bibinfo{number}{5644}) (\bibinfo{year}{2003})
  \bibinfo{pages}{419--422}.

\bibitem[{Stella et~al.(2013)Stella, Zhang, Garc{\'i}a-Vidal, Rubio, and
  Garc{\'i}a-Gonz{\'a}lez}]{Stella:2013}
\bibinfo{author}{L.~Stella}, \bibinfo{author}{P.~Zhang}, \bibinfo{author}{F.~J.
  Garc{\'i}a-Vidal}, \bibinfo{author}{A.~Rubio},
  \bibinfo{author}{P.~Garc{\'i}a-Gonz{\'a}lez}, \bibinfo{title}{Performance of
  Nonlocal Optics When Applied to Plasmonic Nanostructures},
  \bibinfo{journal}{J. Phys. Chem. C}
  \bibinfo{volume}{117}~(\bibinfo{number}{17}) (\bibinfo{year}{2013})
  \bibinfo{pages}{8941--8949}.

\bibitem[{{Teperik} et~al.(2013){Teperik}, {Nordlander}, {Aizpurua}, and
  {Borisov}}]{Teperik:2013}
\bibinfo{author}{T.~V. {Teperik}}, \bibinfo{author}{P.~{Nordlander}},
  \bibinfo{author}{J.~{Aizpurua}}, \bibinfo{author}{A.~G. {Borisov}},
  \bibinfo{title}{Quantum Plasmonics: Nonlocal effects in coupled nanowire
  dimer}, \bibinfo{journal}{Phys. Rev. Lett.}
  \bibinfo{volume}{110}
  (\bibinfo{year}{2013})
  \bibinfo{pages}{[arXiv:1302.3339]}.

\bibitem[{Andersen et~al.(????)Andersen, Jensen, Mortensen, and
  Thygesen}]{Andersen:2013}
\bibinfo{author}{K.~Andersen}, \bibinfo{author}{K.~L. Jensen},
  \bibinfo{author}{N.~A. Mortensen}, \bibinfo{author}{K.~S. Thygesen},
  \bibinfo{title}{Hybridization of quantum plasmon modes in coupled nanowires:
  From the classical to the tunneling regime}, \bibinfo{journal}{Phys. Rev. B}
  \bibinfo{volume}{87}
  (\bibinfo{year}{2013})
  \bibinfo{pages}{[arXiv:1304.4754]}.

\bibitem[{Esteban et~al.(2012)Esteban, Borisov, Nordlander, and
  Aizpurua}]{Esteban:2012}
\bibinfo{author}{R.~Esteban}, \bibinfo{author}{A.~G. Borisov},
  \bibinfo{author}{P.~Nordlander}, \bibinfo{author}{J.~Aizpurua},
  \bibinfo{title}{Bridging quantum and classical plasmonics with a
  quantum-corrected model}, \bibinfo{journal}{Nat. Commun.} \bibinfo{volume}{3}
  (\bibinfo{year}{2012}) \bibinfo{pages}{825}.

\bibitem[{Dong et~al.(2012)Dong, Ma, and Mittra}]{Dong:2012}
\bibinfo{author}{T.~Dong}, \bibinfo{author}{X.~Ma},
  \bibinfo{author}{R.~Mittra}, \bibinfo{title}{Optical response in subnanometer
  gaps due to nonlocal response and quantum tunneling}, \bibinfo{journal}{Appl.
  Phys. Lett.} \bibinfo{volume}{101}~(\bibinfo{number}{23})
  (\bibinfo{year}{2012}) \bibinfo{pages}{233111}.

\bibitem[{Toscano(2013)}]{Toscano:thesis}
\bibinfo{author}{G.~Toscano}, \bibinfo{title}{Semiclassical theory of nonlocal
  plasmonic excitation in metallic nanostructures}, Ph.D. thesis,
  \bibinfo{school}{Department of Photonics Engineering, Technical University of
  Denmark}, \bibinfo{year}{2013}.

\end{thebibliography}

\newpage

\end{document}